\documentclass[aip,rsi,reprint,graphicx]{revtex4-1}
\usepackage{graphicx}
\usepackage{amssymb}
\usepackage{amsmath}
\usepackage{soul}
\usepackage[usenames]{color}

\ifx\pdftexversion\undefined
\usepackage[dvips]{hyperref}
\else
\usepackage{hyperref}
\fi
\hypersetup{
  colorlinks = true, linkcolor = blue
}

\begin{document}

	\title{A large open ratio, time and position sensitive detector for time of flight measurements in UHV.}
	\author{S. Lupone, S. Damoy, A. Husseen, N. Briand, M. Debiossac, S. Tall and P.Roncin} 
	\affiliation{Institut des Sciences Molculaires d’Orsay, CNRS-Universit\'{e} Paris-Sud,Orsay F-91405, France}
	
	\begin{abstract}
		We report on the construction of a UHV compatible 40 mm active diameter detector based on micro channel plates and assembled directly on the feed-throughs of a DN63CF flange. It is based on the charge division technique and uses a standard two inch Si wafer as a collector. The front end electronic is placed directly on the air side of the flange allowing excellent immunity to noise and a very good timing signal with reduced ringing. The important aberrations are corrected empirically providing and absolute positioning accuracy of 500 $\mu$m while a 150 $\mu$m resolution is measured in the center.
	\end{abstract}
	
	\maketitle
	
	\maketitle%
	\section{Introduction}
	Micro channel plate (MCP) based position sensitive detectors are now widely used in atomic and molecular physics as well as in surface science. Such detectors can be place in the focal plane of an electrostatic analyzer  \cite{Brongersma,Roncin86} to allow ion energy measurements or simply placed in view of the interaction zone to record neutral and charged scattered particles scattered \cite{Blavette,Roncin2002} or emitted from the surface\cite{Blavette,Rabalais}. Depending on the respective geometry of the target and primary beam, such detectors are well suited for a variety of surface science techniques such as direct recoil spectroscopy (DRS) low energy ion scattering (LEIS) \cite{Brongersma}, grazing incidence ion scattering \cite{Roncin2002} as well as grazing incidence atom diffraction \cite{Atkinson}.
	
	\section{detector Overview}
\begin{figure}[h] 
		\includegraphics[width=70mm]{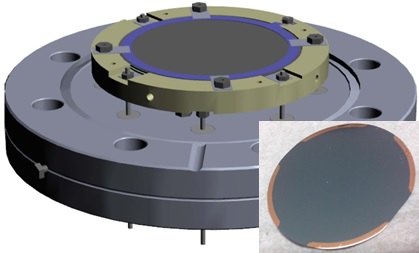}
		\caption{{Designer view of the detector mounted on eight 5 kV UHV feed-throughs theaded on the vacuum side. The inset is a photo of the resistive anode, a two inch silicon wafer with four collection anodes made of evaporated copper.}}
		\label{psd_Scheme}
	\end{figure}

	The detector displayed in figure \ref{psd_Scheme} has been designed to be mounted on a DN63CF port  in an molecular beam epitaxy chamber\cite{Atkinson} where space is limited and UHV compatibility stringent.
	When doing time of flight for DRS or LEIS, the required spatial resolution is only needed  to prevent the broadening of the time of flight peaks due to the rapid variation of the binary collision kinematics with outgoing angle \cite{Rabalais}. With a simple counting detector such as a channeltron the acceptance is usually below a degree opening angle and these aberrations are not considered as a problem \cite{Bauer,esaulov}. Four such detectors were needed justifying a little investment in design. We have used the well-established charge division technique which already showed to be rather compact. Four main characteristics have been developed and will be detailed below. The collector is a standard 2 inch silicon wafer and the detector is directly assembled onto the electrical feed-troughs. The noise sensitive electronic is mounted directly on the air side of the flange taking advantage of modern printed circuit boards facilities which allow a smart combination of electronic and mechanical properties up to the signal and high voltage plugs ({figure \ref{PCB1},  \ref{air_side}}). Finally, the distortion level is larger than anticipated but empirically corrected by a polynomial equation (eq \ref{eq1}) allow a 500 $\mu$m absolute position accuracy.

	\section{the resistive anode}
		
	Delay line detectors \cite{Dowek,David} have the highest count rates and lower dead time but tend to be  large whereas charge division technique have shown to be highly adaptable \cite{Brongersma,Roncin86,Lampton} and robust limiting the need for maintenance. The  distortion-free resistive anodes matching the Gear conditions \cite{Lampton} would have been a obvious choice but the 40 mm active diameter commercially available have a 72 mm overall diameter, too large for DN63CF flange. Since UHV compatibility is mandatory, we have decided to use high impedance 2 inch Si wafers  (high purity single crystal un-doped float zone Si(111), 280 $\mu$m thick one side polished) available at almost no cost from the semi-conductor industry. The size perfectly matches the 40-44 mm active diameter of 50 mm MCP. A simple mask was designed to evaporate four sub micron thick copper collection stripes (fig. \ref{psd_Scheme}) in a high vacuum vessel. 
			
	\section{The vacuum-side assembly} 
\begin{figure}[h]
			\includegraphics[width=55mm]{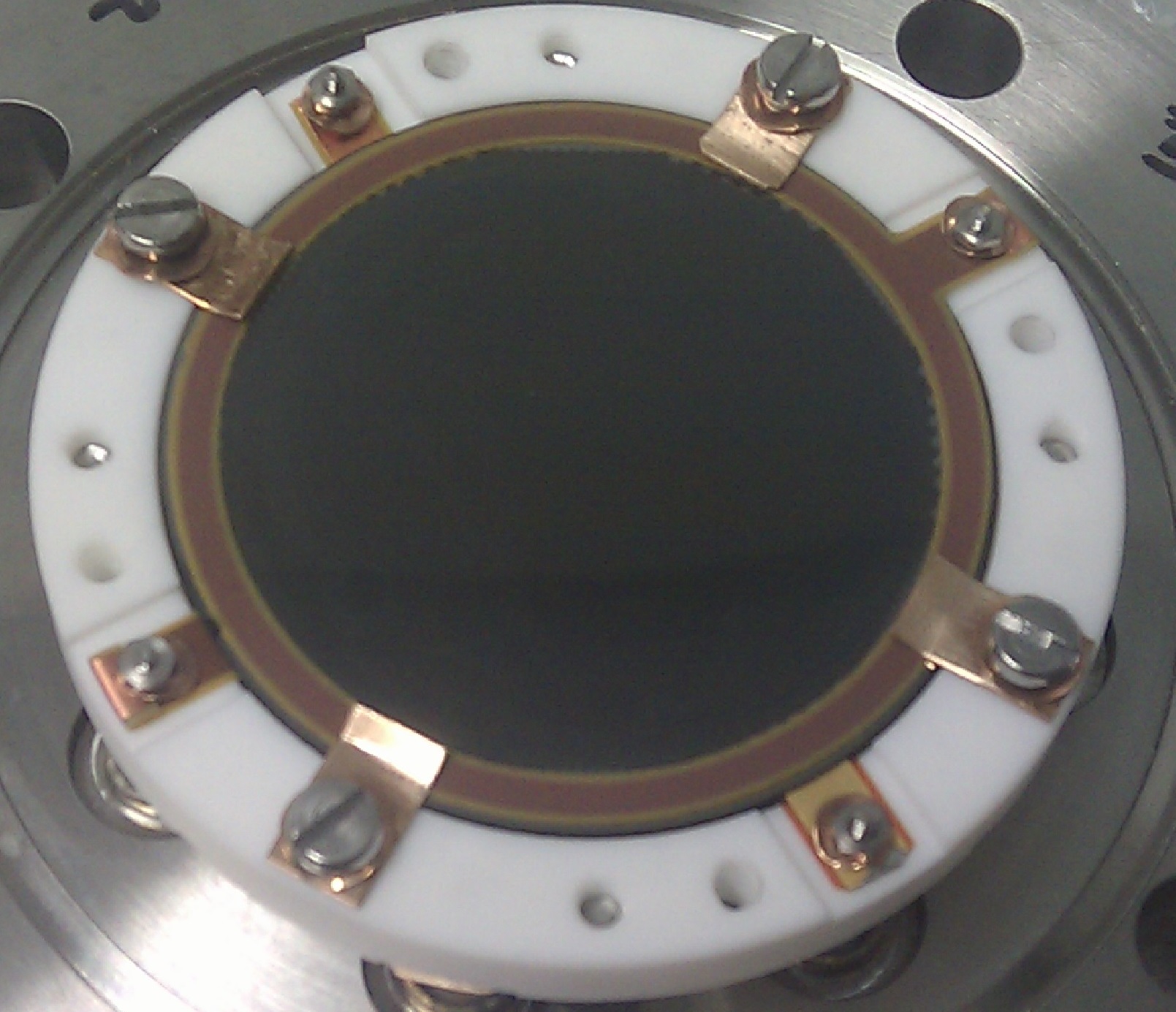}
			\caption{{Vacuum side view of the detector. The four large head screw are only pressing the two MCP stack while the four nuts making electrical contact with copper coated kapton rings hold the detector onto the feed-throughs.}}
			\label{full_view}
		\end{figure}

	The DN63CF flange holding the detector has eight 5 kV feed-troughs welded on a 58 mm diameter. Four are used to bias a two MCP stack and four are used to connect the four copper, charge collection stripes of the anodes. The stems of the feed-throughs on the vacuum side were threaded for 1.4 nuts before welding allowing tight and short UHV connections ({figure \ref{psd_Scheme}}). As shown on figure (fig. \ref{full_view}), the detector structure is a 7 mm thick ring made of Macor or PEEK with the MCP on top and the collector on the bottom side. The detector is sitting on the four anode connectors and compressed by the four other ones connecting the MCP's with copper coated kapton rings. The MCP are compressed separatly by four flat pieces of CuBe acting as a springs, the detector can be assembled separately or directly onto the flange acting as a holder. 
		
	\section{Front end electronic, ADC}
\begin{figure}[h]
			\includegraphics[width=70mm]{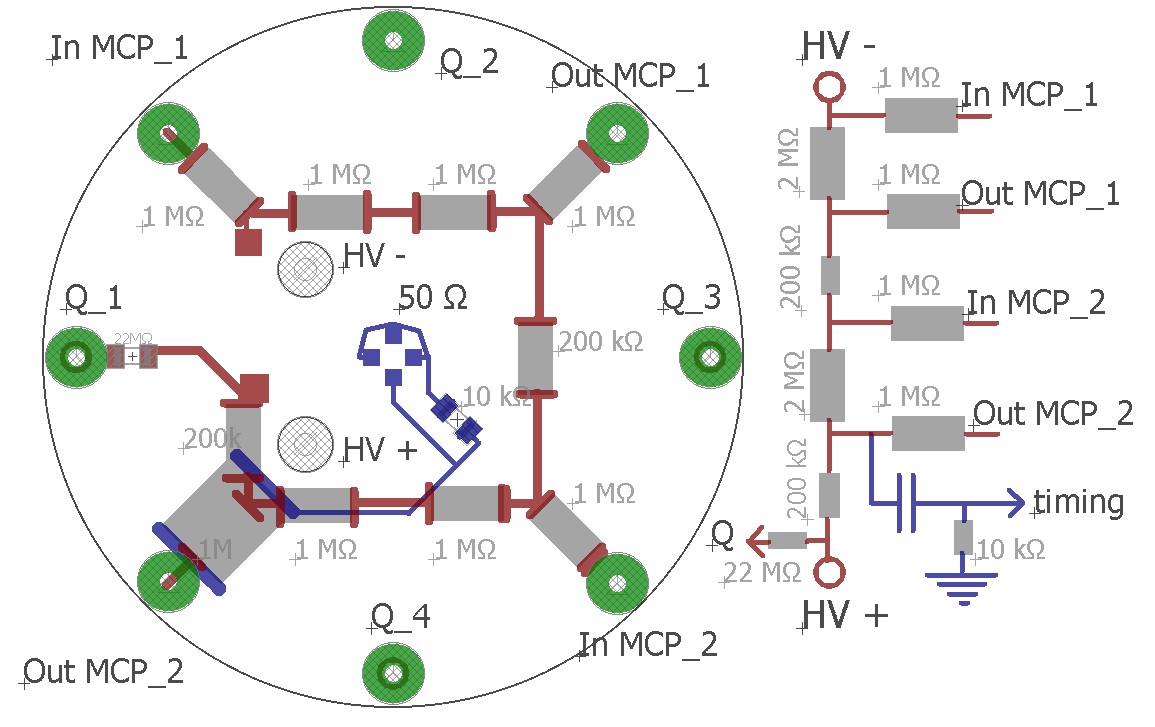}
			\caption{{Designer view of the base PCB with the HV divider made of surface mount 1kV resistors. It includes MCP protection resistors, inter-plate acceleration and acceleration to the collector. The electrical connection is sketched on the right.}}
			\label{PCB1}
		\end{figure}

\begin{figure}[h]

		\includegraphics[width=65mm]{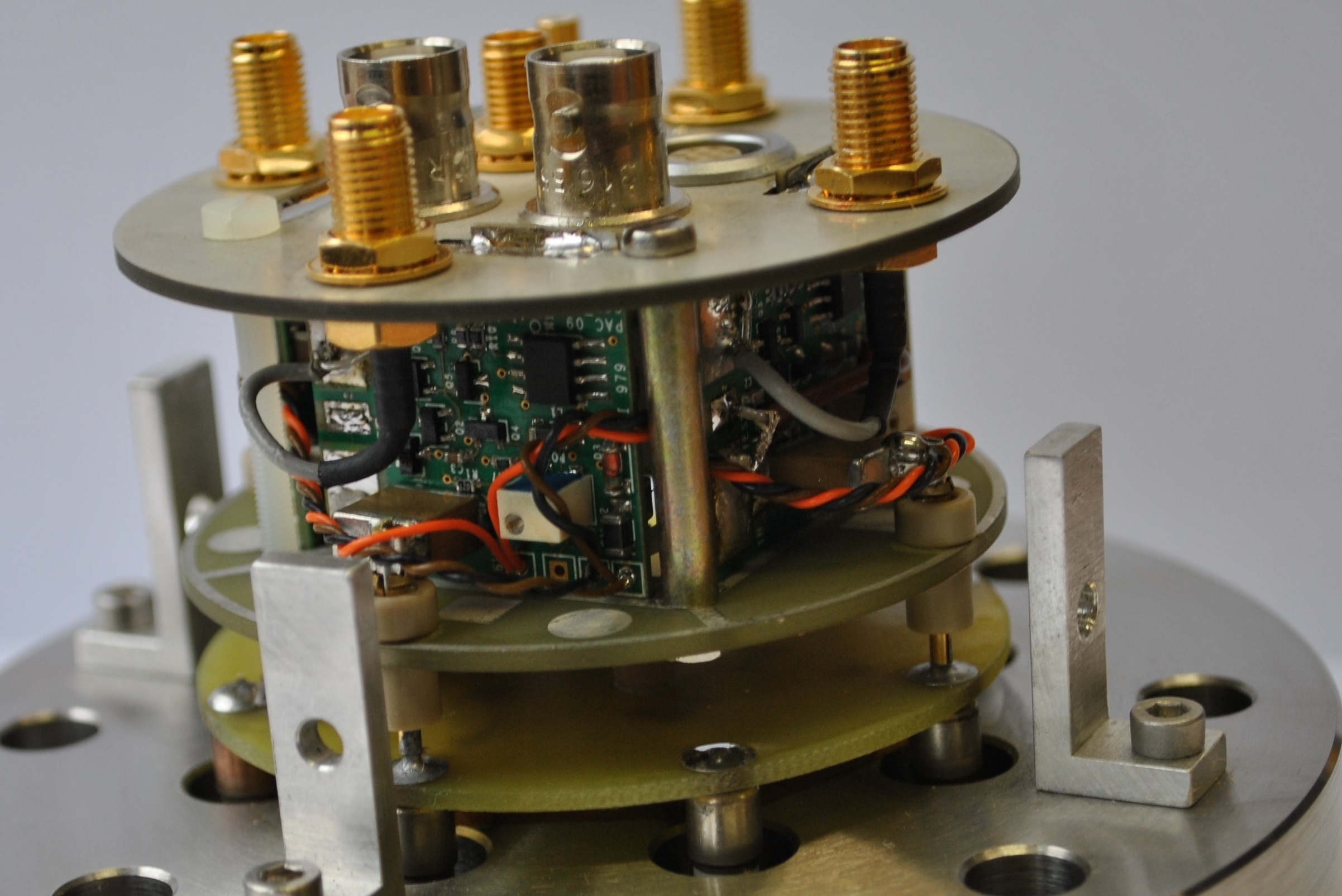}
		\caption{{Air side views of the front end electronic with the three PCBs holding the voltage divider, the charge preamplifier and all the connectors}}
		\label{air_side}
	\end{figure}

	The detector is designed to operate indifferently for neutrals, positive or negative ions (or electrons), provided that neither HV$^+$ nor HV$^-$ exceed 5 kV  while HV$^+$-HV$^-$ $\lesssim$ 2kV. The decoupling and the readout electronic are placed close to the detector. This is easily achieved with modern printed circuit board (PCB) facilities offering custom machining allow mechanical aspects to be part of the design. The assembly is made of three circular PCB. The first one (fig.\ref{PCB1} ) holds the high voltage divider to bias the MCP's as well as the decoupling of the fast timing signal from the output of second MCP. This PCB is soldered to the flange by the four MCP feed-throughs. 
	The four other ones, connected to the anode, are capped by the tip of a 1.55 mm ITT CANNON connector tip and used as male connectors for the 2$^{nd}$ and 3$^{rd}$ PCB (see fig.\ref{air_side}). The 2$^{nd}$ PCB imprisons the female connector in a plastic cylinder. The PAC09 charge preamplifier purchased at the Orsay institute of nuclear physics (IPNO) are connected to these pin via four 5 kV, 2 nF ceramic capacitor. These are soldered to 3$^{rd}$ PCB and compressed between the 2$^{nd}$ PCB and 3$^{rd}$ PCB. The third PCB holds all the connectors; two BNC high voltage connectors to bias MCP's, four SMA connectors for preamplifier outputs, a test input common to all preamplifiers and a SMA output for the fast timing signal. No connection to the flange voltage is present yet. This specificity could be interesting for specific purpose but not in our case. The ground connection is achieved by a simple metal tubing fixed directly onto the flange. The contact with the top PCB turned out to be more tricky than expected and CuBe springs pressing the top PCB have been needed. 
	
	The rest of the electronics is rather standard with four triangular shape linear amplifiers with a 2 $\mu$s shaping constant (also from IPNO). The analog to digital conversion is achieved with low cost NI6143, eight channels, 250 kHz simultaneous sampling ADC. The maximum count rate should therefor lie in the 100 kHz range. Higher count rate would require faster digitizer but would rapidly hit the pileup limit of the amplifiers. The cards offers an embedded 50 ns counter which could allow stand alone coincidence operation of two PSD. For better timing resolution, an additional electronics is needed but could be as simple as a time to amplitude converter feed in one of the four unused channels. The fast signal decoupled from second MCP is probably in the 1 ns range but could not be measured accuratly on our test system. 
	
	\section{Calibration and aberration correction} 
\begin{figure}[h]
		\includegraphics[width=\columnwidth]{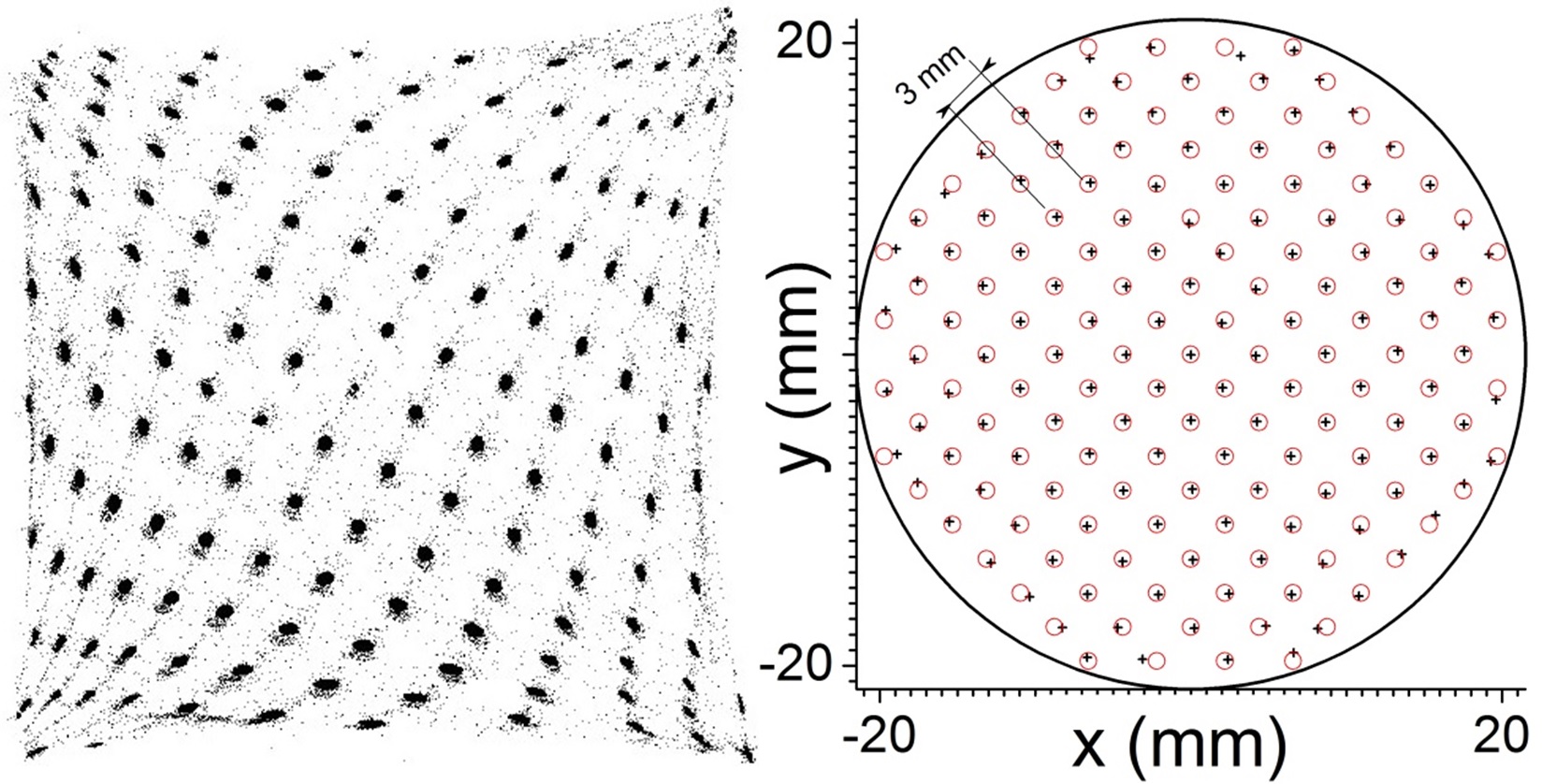}
		\caption{{Left is a raw image of all impact through the mask. After correction by formula \ref{eq2}, the image on the right shows that most of the spots centers are within 500 $\mu$m of their ideal position as illustrated by the 1 mm circles sitting at mask locations. The 43 mm circle is only to guide the eyes.}}
		\label{psd_corrections}
	\end{figure}

	To calibrate the detector, a copper mask with an array of 0.5 mm holes every 3 mm is mounted on the detector entrance.The large size is to allow fast acquisition from all the holes when illuminated by a 2 keV Ar$^+$ beam. At variance, the central hole is only 0.1 mm in diameter allowing resolution information. The image in figure \ref{psd_corrections} is obtained with the usual formula (eq. \ref{eq1}) where Q$_i$ are the four charges. The image displays all ion impact illuminating 145 holes and shows very important distortions. The measured size of the central spot imaging the 100 $\mu$m hole is 200 $\mu$m fwhm suggesting an intrinsic resolution of 150 $\mu$m fwhm. 
		\begin{equation}
		\ X = \frac{Q_{left} -  Q_{right}}{2\sum_i Q_i}, Y =\frac{ Q_{botom} -  Q_{top}}{2\sum_i Q_i}
		\label{eq1}
		\end{equation}	
	To recover a correct image, we have used brut force; each observed spot is fitted by a 2D gaussian and is associated to integer indexes i,j, then a 4$^{th}$ order polynomial (eq. \ref{eq2}) is optimized to adjust X, Y observed values to X', Y' values sitting on the mask position (fig.\ref{psd_corrections}).

	\begin{equation}
		\ X' = \sum_i \sum_j a_{mn} X^m Y^n, Y' = \sum_i \sum_j b_{mn} X^m Y^n 
		\label{eq2}
	\end{equation}
	
	
	\section{critical analysis}
 The incredible quality of the mono-cristalline Si Wafer did not turn out to be such a smart choice. The UHV compatibility is perfect but the level of distortion is higher than with our two circular test anode made of evaporated amorphous Ge or of simple painted carbon film where a single parameter a$_{11}$ (a$_{11}$.XY) identical for X and Y direction (a$_{11}$=b$_{11}$) was enough to correct most distortions in the central region. Since copper film on kapton foil are easy to purchase directly from PCB industry, simple carbon or Ge evaporation could be a better choice. The threaded feed-throughs are excellent to compress the detector but those connecting the anode and used in compression would probably be easier to adjust with, for instance, male twist pins crimped on the stems and inserted into perforated screw. When additional space is available, a fourth PCB layer hosting linear amplifiers could suppress the need of bulky NIM crate. Surface mount High voltage resistor was a success but the use high voltage capacitors with X7R technology (visible on fig. \ref{air_side}) should be avoided, indeed, the insulating resistance in the 10-100 M$\Omega$ range is too low so that, when operated at high voltage, the collector injects charges directly into the preamplifier which then display a significant noise. The trick of capping the stem with a connector tip was adopted after the 1st PCB was soldered, if applied also to the 1$^{st}$ PCB connection all PCB could be removed making outside final impedance check-list and bake-out more simple. The present design is rather conservative, the outer diameter is 64 mm which fits easily in the 70 mm of so called large bore 63CF flanges which are standard on UHV gate valve and available at most UHV providers.

	\section{Acknowledment}
	We are grateful to O. Jagutzki from Roentdek Gmbh for kindly giving us a circular glass anode with evaporated Ge to test our approach. M. Albaret is acknowledged for his bright ideas combining electronic and mechanic,  A. Momeni for his continuous support and X. Fl\'{e}chard for help in the correction algorithm. This work has been supported by Agence Nationale de la Recherche ANR-07-BLAN-0160.


\end{document}